\documentclass[10pt, conference, compsocconf]{IEEEtran}
\usepackage{url}

\usepackage{color}
\usepackage{wrapfig}
\usepackage{amsmath,amsthm,amssymb}
\usepackage{graphicx,algorithm,algorithmic}
\usepackage{subcaption}
\usepackage{url}
\usepackage{paralist}
\usepackage{xspace}

\usepackage{tikz}
\usetikzlibrary{positioning,patterns,arrows,shapes,calc}
\usepackage{pgfplots,pgfplotstable}
\usepgfplotslibrary{external}
\usepackage{etex}
\usepackage{array}
\let\labelindent\relax
\usepackage{enumitem}
\usepackage{mdwtab}
\usepackage{booktabs}
\usepackage{dcolumn}

\usepackage{listings}
\usepackage{alltt, textcomp, color, verbatim}
\lstdefinestyle{cpp} {
		language=C++,                        % choose the language of the code
		basicstyle=\ttfamily\footnotesize,   % the size of the fonts that are used for the code
		tabsize=2,                           % sets default tabsize to 2 spaces
		showstringspaces=false,              % show spaces adding particular underscores
		showtabs=false,                      % show tabs within strings adding particular underscores
		keywordstyle=\color{blue},           % keyword style
		identifierstyle=\color{black},       % identifier style
		emphstyle=\color{black}\bf,          % emphasis style
		commentstyle=\color{gray}\slshape,   % comment style
		stringstyle=\color{gray},            % string literal style
		aboveskip=\baselineskip,             % skip space when starting code environment
		xleftmargin=10pt, xrightmargin=10pt, % code margins
		frame=lines,                         % adds a frame around the code
		numbers=left,                        % where to put the line-numbers
		numberstyle=\tiny,                   % the size of the fonts that are used for the line-numbers
		numbersep=10pt,                      % how far the line-numbers are from the code
}
\lstnewenvironment{cpp}[1][]{%
   \noindent
   \minipage{\linewidth}
  \lstset{%
	style=cpp,
    #1
  }%
}{\endminipage}%

\newcommand{\fixme}[1]{{\bf \color{red}{#1}}}
\newcommand{\COMMENTOUT}[1]{}

\newcommand{\ie}{\textit{i.e.}\xspace}
\newcommand{\etal}{\textit{et~al}\xspace}
\newcommand{\DOT}{\textsc{Dot}\xspace}
\newcommand{\GER}{\textsc{Ger}\xspace}

\newcommand{\GEMV}{\textsc{Gemv}\xspace}

\newcommand{\GEMM}{\textsc{Gemm}\xspace}
\newcommand{\GEMMs}{\textsc{Gemm}s\xspace}
\newcommand{\BGEMM}{\textsc{BatchedGemm}\xspace}
\newcommand{\BGEMV}{\textsc{BatchedGemv}\xspace}

\newcommand{\SBGEMM}{\textsc{StridedBatchedGemm}\xspace}

\newcommand{\MKL}{\textsc{Mkl}\xspace}
\newcommand{\CUBLAS}{\textsc{CuBlas}\xspace}

\newcommand{\BLAS}{\textsc{Blas}\xspace}
\newcommand{\BLIS}{\textsc{Blis}\xspace}
\newcommand{\BTAS}{\textsc{Btas}\xspace}
\newcommand{\CYCLOPS}{\textsc{Cyclops}\xspace}

\newcommand{\abs}[1]{\left| {#1} \right|}

\renewcommand{\v}{\mathbf}

\newtheorem{lemma}{Lemma}
\newtheorem{corollary}{Corollary}

\ifCLASSINFOpdf
\else
\fi
\begin{document}
\title{Tensor Contractions with Extended BLAS Kernels\\ on CPU and GPU}


\author{\IEEEauthorblockN{Yang Shi \IEEEauthorrefmark{1}, U. N. Niranjan \IEEEauthorrefmark{2}, Animashree Anandkumar \IEEEauthorrefmark{1}}
\IEEEauthorblockA{\IEEEauthorrefmark{1} EECS Department, \IEEEauthorrefmark{2} ICS Department\\
University of California, Irvine\\
Irvine, USA\\
Email: \{shiy4,un.niranjan,a.anandkumar\}@uci.edu}
\and
\IEEEauthorblockN{Cris Cecka}
\IEEEauthorblockA{NVIDIA Research\\
Santa Clara, USA\\
Email: criscecka@gmail.com}
}


%Telephone: (800) 555--1212, Fax: (888) 555--1212}


\maketitle

\begin{abstract}

Tensor contractions constitute a key computational ingredient of numerical multi-linear algebra. However, as the order and dimension of tensors grow, the time and space complexities of tensor-based computations grow quickly.  In this paper, we propose and evaluate new \BLAS-like primitives that are capable of performing a wide range of tensor contractions on CPU and GPU efficiently. We begin by focusing on single-index contractions involving all the possible configurations of second-order and third-order tensors. Then, we discuss extensions to more general cases.

% We present the natural starting point, namely,   single-index contractions of a third-order tensor, in detail and discuss extensions to more general cases.

Existing approaches for tensor contractions spend large amounts of time restructuring the data which typically involves explicit copy and transpose operations. In this work, we summarize existing approaches and present library-based approaches that avoid memory movement. Through systematic benchmarking, we demonstrate that our approach can achieve 10x speedup on a K40c GPU and 2x speedup on dual-socket Haswell-EP CPUs, using \MKL and \CUBLAS respectively, for small and moderate tensor sizes. This is relevant in many machine learning applications such as deep learning, where tensor sizes tend to be small, but require numerous tensor contraction operations to be performed successively. Concretely, we implement a Tucker decomposition and show that using our kernels yields atleast an order of magnitude speedup as compared to state-of-the-art libraries.

\end{abstract}


% \fi
\IEEEpeerreviewmaketitle

%=======================================================================

%=======================================================================
\section{Introduction and Scope}
\label{sec:introduction}
\noindent {\em Multilinear algebraic} computations, are ubiquitous in multiple scientific domains such as machine learning and modern data science~\cite{Anima:2014:JMLR}, quantum chemistry and physics~\cite{khoromskaia2015tensor}, signal and image processing~\cite{goyal2014fourier}, chemometrics~\cite{bro2003new}, and biochemistry~\cite{kazeev2013direct}. The study of tensor computations has a long and diverse history, as early as in the work by Hitchcock~\cite{hitchcock1927expression}. The domains and references provided herein are by no means exhaustive but merely a small representative sample of the various flavors in which tensor computations are used in science. \textit{Tensors} are multi-way arrays which can be viewed as a generalization of matrices to allow \textit{multi-modality} in data. Tensor \textit{contractions} play a central role in a variety of algorithms and applications; for a motivating example, see Section~\ref{sec:ALS}. However, non-trivial performance bottlenecks in several application areas are encountered due to the high space and time complexities associated with tensor computations. In this paper, motivated by the recent increased interest from machine learning and deep learning, we propose and study library-based communication avoiding approaches for performing tensor contractions.

Conventional approaches for computing general tensor contractions rely on {\em matricization}, the logical or explicit restructuring of the data so that the computation can be performed with a sequence of Basic Linear Algebra Subroutine (\BLAS) library calls. The \BLAS routines provide efficient and portable implementations of linear algebra primitives, with many fast implementations existing across many architectures~\cite{Aydin:2011:HPCA}. %Thus, the computation may be efficient, but the restructuring is almost always implemented with explicit out-of-place copies or transpositions, see Section~\ref{sec:btas}.

To this point, the GEneral Matrix Multiply (\GEMM) primitive specified within the \BLAS library is possibly the most optimized and widely used routine in scientific computing. Noting that the basic theoretical computational and communication complexities of most tensor contractions is equivalent to that of \GEMM, these computations should scale equally well. However, we find that existing tensor libraries such as the \textsc{Tensor Toolbox} and \textsc{Cyclops Tensor Framework} perform explicit data transposition to compute almost all tensor contractions and the cost of data restructuring often dominates the cost of the actual computation. %In Section~\ref{sec:motivation}, we show that more than 50$\%$ of the total evaluation time can be taken up in restructuring the data, especially for small tensor contractions. 
Other approaches have previously proposed intrusive compiler and static analysis solutions, whereas we provide a much simpler library-based solution~\cite{Li:2015:TensorMatrixMult,Lu:2012:EPM}.

%This is a significant overhead in many machine learning applications such as deep learning since one encounters a large number of contractions over small tensors~\cite{Richard:2013:EMNLP}.

\begin{comment}
In this paper, we propose extensions to \BLAS that enable in-place evaluation of general tensor contractions with the potential to achieve performance comparable to \GEMM.
\end{comment}
\paragraph{Findings and contributions}

We introduce a new \BLAS primitive, known as \SBGEMM, that allows the majority of tensor contractions to be computed without any explicit memory motion. We detail the so-called exceptional cases that cannot be evaluated with \SBGEMM and demonstrate that an efficient solution exists with another small extension to the primitive.

We demonstrate performance improvement using our approach on both CPU and GPU in direct benchmarks in addition to an application study. The Tucker decomposition is an important tensor application in machine learning wherein the advantage of our strategy compared to existing libraries is clear. 

Finally, the value of this approach and its applications are being recognized by NVIDIA. As of this writing, the proposed interface exists in the \CUBLAS 8.0 Release Candidate and is likely to appear the official release later this summer.

%However, the true benefit of our approach is over and beyond this speedup. We have proposed a simplified interface for solving a large class of tensor contractions, and this provides opportunities to optimize the new \BLAS primitives on different platforms. Moreover, our proposed approach will have real-world impact very soon: we have it in high confidence that NVIDIA will be releasing a implementation of \SBGEMM with \CUBLAS 8.0.

\vspace*{-5pt}
%%=======================================================================
\section{Background}
\label{sec:background}
\vspace*{-5pt}
\subsection{Related Work}
\label{sec:related}

Peise \etal~\cite{Peise:2015:TensorBLAS} extended results from Napoli \etal~\cite{Napoli:2014:EfficientBLAS} in mapping tensor contractions to sequences of \BLAS routines and modeling the performance of these mappings. In this work, they systematically enumerate and benchmark combinations of possible \BLAS kernels one could use to compute a given tensor contraction to conclude that the best performing algorithms involve the \GEMM kernel. Some evaluation strategies are neglected to be considered, such as \textit{flattening} or developing new, generic linear algebraic subroutines that could yield improved performance. %Furthermore, there is no discussion regarding performance on GPUs and very little parallel consideration.

%Lu \etal~\cite{Lu:2012:EPM} produce an optimizing compiler to determine the number and sequence of transpositions of a general tensor contraction so that the evaluation can be performed with a \GEMM call. However, they do not consider batched operations such as \BGEMM. %Many transpose operations are required. The system produces many intermediates and also neglects non-uniform striding of the leading dimension. \fixme{Revisit [Cris]}
%Lu \etal~\cite{Lu:2012:EPM} produce an optimizing compiler to determine the number and sequence of transpositions of a general tensor contraction so that the evaluation can be performed with a \GEMM call ... TCE-targeted. Doesn't consider batched GEMM: Produces many intermediates, Many required transpose operations. Also neglects non-uniform lda striding... \fixme{[Cris]}

Li \etal~\cite{Li:2015:TensorMatrixMult} also recognizes the cost of explicit copies and proposes evaluation strategies exactly comparable to the flattening and batching strategies addressed in this paper. Their discussion of {\em loop modes} and {\em component modes} map to our discussion of {\em batch modes} and {\em \GEMM modes}. However, Li \etal do not discuss strategies beyond tensor-times-matrix multiply. Furthermore, they only consider {\em mode-$n$} tensor-times-matrix contractions of the form $Y_{i_1\cdots i_{n-1} j \cdots i_N} = \sum_{i_n} X_{i_1 \cdots i_N} U_{j i_n}$, which avoids the more complicated cases in this paper. Abdelfattah \etal~\cite{abd2016iccs} presents a framework using batched \GEMM for tensor contractions on GPUs. However, they focus on optimizing only limited number of tensor contraction kernels on extreme small size tensors. Other works in \cite{allam2006ipdps} \cite{nelson2015icpp} improve the tensor computation performance by doing loop reorganization and fusion.

%Furthermore, our proposed notation makes the discussion much easier to follow. Ideally, the heuristics to determine the optimal evaluation configuration presented by Li \etal could be adapted to use the more efficient and portable primitives proposed in this paper for general in-place tensor-times-tensor multiply algorithms.
The \SBGEMM interface proposed in this paper has previously been mentioned by Jhurani \etal~\cite{Jhurani:2015:GEMMInterface} as a low-overhead interface for multiple small matrices on NVIDIA GPUs. Jhurani proposes the same interface for \CUBLAS that we propose in this paper and focuses on implementation concerns. In this work, we treat \SBGEMM as an available primitive, benchmark evaluation strategies that utilize it, and examine how it may be further extended for use in multi-linear algebra.

The \BLAS-like Library Instantiation Software (\BLIS) framework~\cite{VanZee:2015} offers \GEMMs which support non-unit strides in {\em both} the row and column dimensions, which are attractive solutions to some of the problems in this paper. However, performance is expected to suffer due to decreases in cache line utilization, %prefetch bandwidth, 
and SIMD opportunities.

Recent improvements in parallel and distributed computing systems have made complex tensor computation feasible. TensorFlow~\cite{tensorflow2015-whitepaper} can handle multi-linear algebra operations and it is primarily a data-flow and task-scheduling framework for machine learning. 
%also proposed tensor contraction operations in the framework. However it is not accessible from Python API yet. They may add the operations in TensorFlow package or from 3rd party libraries. As they point out from their tutorial, calling efficient computation libraries outside of Python is not optimal because there can be lots of overhead from switching back to Python. \fixme{Revisit paragraph. Python shouldn't be the issue.}

%In general, it appears that available libraries for tensor algebra under-utilize available \BLAS libraries.% \fixme{[Cris] Overview}
\vspace{-5pt}
\subsection{Notation}
\label{sec:notation}

 We denote tensors by uppercase letters, indices by lowercase letters and index lists by calligraphic letters. We assume all indexing is zero-based. $\mathbb{R}$ denotes the set of real numbers.

% \begin{definition}
The \textbf{order} of a tensor is the number of \textbf{modes} it admits.
%\end{definition}
A scalar is a zeroth-order tensor, a vector is a first-order tensor, a matrix (say $A_{mn}$) is a second-order tensor with the rows (indexed by $m$) being the first mode and columns (indexed by $n$) being the second mode, and a three-way array (say $A_{mnp}$) is a third-order tensor with the first, second and third modes indexed by $m$, $n$, and $p$, respectively. Note that we use the term \textit{index} to name a mode and iterate through the elements in that mode.%\fixme{I don't know about the formal definition structure here... I like the more informal version we had before.}

%\begin{definition}
The \textbf{dimension} of the $i^\text{th}$ mode, denoted {\tt dim<$i$>}, is the number of elements it contains.
%\end{definition}
The dimension of a mode of a tensor is denoted by the bold lowercase letter of the respective index; for example, the third-order tensor $A_{mnp}$ has dimension {\tt dim<$0$>}$ \times ${\tt dim<$1$>}$ \times ${\tt dim<$2$>} or $\mathbf{m} \times \mathbf{n} \times \mathbf{p}$ where the first mode (indexed by $m$) takes values $0, \ldots, \mathbf{m}-1$, the second mode (indexed by $n$) takes values $0, \ldots, \mathbf{n}-1$, the third mode (indexed by $p$) takes values $0, \ldots, \mathbf{p}-1$. %\fixme{We're using capitol letters rather than bold lowercase elsewhere... Change to capitol?}

%In this section, we detail our notation to map tensor-matrix contractions to their corresponding \BLAS-like computations.

We follow Einstein summation convention to represent tensor contractions.%, i.e., summation is implicitly understood to occur among the common indices of the involved operands. 
 A general tensor contraction is written as
\begin{align}
C_{\mathcal{C}} = \alpha \, A_{\mathcal{A}} \, B_{\mathcal{B}} + \beta \, C_\mathcal{C}
\label{eqn:GETM}
\end{align}
where $\mathcal{A}, \mathcal{B}, \mathcal{C}$ are ordered sequences of indices such that %. Using notation from set theory, note that the index sets must satisfy 
%\begin{align}
$\mathcal{C} \equiv (\mathcal{A} \cup \mathcal{B})  \setminus (\mathcal{A} \cap \mathcal{B})$. 
%\label{eqn:set_diff}
%\end{align}
%for the contraction to be well-defined.
The indices in $\mathcal{A} \cap \mathcal{B}$ are called \textit{contracted indices}. The indices in $\mathcal{C}$ are called \textit{free indices}.
\vspace{-5pt}
\subsection{An Important Practical Application}
\label{sec:ALS}
\vspace{-2pt}
In unsupervised learning, tensor decomposition~\cite{Anima:2014:JMLR} is gaining a lot of attention and is the crux of model estimation via the method of moments. A variety of problems such as topic model estimation, Gaussian mixtures model estimation,  and social network learning can be provably, consistently and efficiently solved via the tensor decomposition techniques under certain mild assumptions.

The basic building blocks of these algorithms involve tensor contractions. Two frequently used tensor decomposition methods are the CP decomposition~\cite{Harshman:1970:UCLA} and the Tucker decomposition~\cite{Tucker:1966:Psy}. In~\cite{Alex:2002:ECCV}, the authors 
%consider each mode of the tensor as people's face with different head poses, expressions and lighting conditions. By
use the Tucker decomposition to extract new representations of the face images %which map all images of a person to a vector, 
despite different expressions or camera viewpoints. To illustrate the fundamental importance of tensor contractions, we will pick one of the most common tensor decomposition algorithms, namely the higher-order orthogonal iteration (HOOI)~\cite{lde2000siam} for asymmetric Tucker decomposition, and use it as a case-study. In the Einstein notation, the factorization of a third-order tensor $T \in \mathbb{R}^{\mathbf{m} \times \mathbf{n} \times \mathbf{p}}$ is given by $T_{mnp} = G_{ijk}A_{mi}B_{nj}C_{pk}$, where $G \in \mathbb{R}^{\mathbf{i} \times \mathbf{j} \times \mathbf{k}}$ is the core tensor, $A \in \mathbb{R}^{\mathbf{m} \times \mathbf{i}}$, $B \in \mathbb{R}^{\mathbf{n} \times \mathbf{j}}$, $C \in \mathbb{R}^{\mathbf{p} \times \mathbf{k}}$. %Elementwise, $T_{ijk} = \sum_{p = 1}^{m_1} \sum_{q = 1}^{m_2}  \sum_{r = 1}^{m_3} G_{pqr}A_{ip}B_{jq}C_{kr}$.
From Kolda et al~\cite{Kolda:2009:SIAM}, we summarize the algorithm for the third-order tensor case in Algorithm~\ref{alg:als}. Following their notation, $T_{(r)}$ denotes the mode-$r$ unfolding of tensor $T$. For further technical details, we refer the reader to Kolda et al~\cite{Kolda:2009:SIAM}.

\begin{algorithm}[t]
   \algsetup{linenosize=\footnotesize}
  \small
\begin{algorithmic}[1]
\REQUIRE Tensor $T \in \mathbb{R}^{\mathbf{m} \times \mathbf{n} \times \mathbf{p}}$, core tensor size $\mathbf{i}$, $\mathbf{j}$, $\mathbf{k}$, number of  iterations $\mathcal{T}$.
\ENSURE Factors $A^{\mathcal{T}}$, $B^{\mathcal{T}}$, $C^{\mathcal{T}}$ and core tensor $G$
\hrule
\STATE Set $t=0$; 
\STATE Initialize $A^0 \leftarrow \mathbf{i}$ leading left singular vector of $T_{(1)}$ \\ $\quad \quad \quad$ $B^0 \leftarrow \mathbf{j}$ leading left singular vector of $T_{(2)}$\\$\quad \quad \quad$ $C^0 \leftarrow \mathbf{k}$ leading left singular vector of $T_{(3)}$
\WHILE {$t < \mathcal{T}$}
\STATE $Y_{mjk} = T_{mnp} B_{nj}^{t} C^{t}_{pk}$
\STATE $A^{t+1}$ $\leftarrow$ $\mathbf{i}$ leading left singular vector of $Y_{(1)}$.
\STATE $Y_{ink} = T_{mnp} A^{t+1}_{mi} C^{t}_{pk}$
\STATE $B^{t+1}$ $\leftarrow$ $\mathbf{j}$ leading left singular vector of $Y_{(2)}$.
\STATE $Y_{ijp} = T_{mnp} B^{t+1}_{nj}  A^{t+1}_{mi}$
\STATE $C^{t+1}$ $\leftarrow$ $\mathbf{k}$ leading left singular vector of $Y_{(3)}$.
\ENDWHILE
%\UNTIL $t < \mathcal{T}$.
\STATE $G_{ijk} = T_{mnp} A^{\mathcal{T}}_{mi} B^{\mathcal{T}}_{nj}  C^{\mathcal{T}}_{pk}$
\end{algorithmic}
\caption{Tucker decomposition algorithm.}
\label{alg:als}
\end{algorithm}

%Referring to Table~\ref{tab:contract}, the tensor contractions that we encounter in this algorithm are in step 4, 6, 8 and step 11.
\begin{comment}
\begin{compactitem}
\item \textit{Step 4: }$C_{mnp} = A_{kn}B_{mkp}$, $C_{mnp} = A_{kp} B_{mnk}$.
\item \textit{Step 6: }$C_{mnp} = A_{km} B_{knp}$, $C_{mnp} = A_{km} B_{knp}$.
\item \textit{Step 8: }$C_{mnp} = A_{km} B_{knp}$, $C_{mnp} = A_{kn} B_{mkp}$.
\item \textit{Step 11: }$C_{mnp} = A_{km} B_{knp}$, $C_{mnp} = A_{kn} B_{mkp}$, $C_{mnp} = A_{kp} B_{mnk}$.
\end{compactitem}\end{comment}
%\fixme{Can we make Algorithm 1 smaller?}
\vspace{-5pt}
\subsection{Conventional Tensor Contraction}
\label{sec:btas}

%\paragraph{BTAS}
%BTAS~\footnote{Available at \url{http://itensor.org/btas/}} is a high-level library written in C++ for performing tensor contractions. BTAS considers arbitrary tensor operands of any order and dimension and multi-index tensor contraction operations are allowed, whenever the contracted indices are dimension-compatible. This is accomplished in four stages: 
The conventional approach for tensor contraction is to {\em matricize} the tensors via transpositions and copies. Libraries such as Basic Tensor Algebra Subroutines (\BTAS)~\cite{BTAS_Software}, MATLAB Tensor Toolbox~\cite{TTB_Software,TTB_Dense}, and Cyclops Tensor Framework~\cite{solomonik2013cyclops} all perform some version of matricization, which is typically performed in four steps:
\begin{compactenum}
\item Consider a general tensor contraction of the form \eqref{eqn:GETM}. Define the index sets $\mathcal{K}$, $\mathcal{I}$, $\mathcal{J}$ as
\vspace{-0.3em}
\begin{align*}
\mathcal{K} = \mathcal{A} \cap \mathcal{B}, \quad
\mathcal{I} = \mathcal{A} \setminus (\mathcal{A} \cap \mathcal{B}), \quad
\mathcal{J} = \mathcal{B} \setminus (\mathcal{A} \cap \mathcal{B})
\end{align*}
\vspace*{-1.25em}
\item Permute tensors $A$, $B$, and $C$ into the form
\begin{align}
C_{\mathcal{I}\mathcal{J}} = \alpha \, A_{\mathcal{I}\mathcal{K}} \, B_{\mathcal{K}\mathcal{J}} + \beta \, C_{\mathcal{I}\mathcal{J}}
\label{eqn:BTASeq}
\end{align}
\vspace*{-10pt}
\item Evaluate \eqref{eqn:BTASeq} using one of four BLAS kernels:
\begin{align*}
\begin{cases}
\DOT  & \abs{\mathcal{K}} = \abs{\mathcal{A}} \text{ and } \abs{\mathcal{K}} = \abs{\mathcal{B}} \\
\GER  & \abs{\mathcal{K}} = 0 \\
\GEMV & \abs{\mathcal{K}} = \abs{\mathcal{A}} \text{ xor } \abs{\mathcal{K}} = \abs{\mathcal{B}} \\
\GEMM & \text{else}
\end{cases}
\end{align*}
\vspace*{-5pt}
\item Permute the result, $C_{\mathcal{I}\mathcal{J}}$, into the desired output, $C_{\mathcal{C}}$.
\end{compactenum}
This approach to tensor contractions is completely general -- it works for any two tensors of arbitrary order and any number of contraction indices. However, for even the simplest contractions, the cost of explicitly permuting the tensor data typically outweigh the cost of the computation to be performed. See Section~\ref{sec:motivation} for examples.

%Iterative improvements to the above algorithm can be made by batching \BLAS calls to avoid the explicit permutations. Napoli and Peise \etal~\cite{Napoli:2014:EfficientBLAS,Peise:2015:TensorBLAS} enumerate and model the performance of all possible such calls. 
%In essence, though this is a general approach, \BTAS makes no attempt to exploit the leading dimension for flattening, minimize the number of explicit memory operations, or use batched \BLAS operations.

%=======================================================================
% eof

%%=======================================================================
%%\section{Tensor Decomposition and Applications}
%%\label{sec:applications}
%%\input{applications}
%
%%=======================================================================
\section{Approach}
\label{sec:approach}
%While the algorithm presented in Section~\ref{sec:btas} is general and applicable to any tensor contraction, the cost of explicit copy and/or transposition is significant. 
In this section, we present library-based evaluation strategies for performing general tensor contractions in-place -- without explicit copies and/or transpositions.
\vspace{-5pt}
\subsection{Motivating Observations}
\label{sec:motivation}

\begin{figure}[ht]
\centering
\begin{subfigure}[t]{0.45\linewidth}
\includegraphics[]{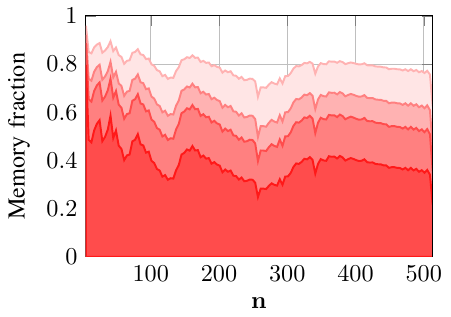}
\end{subfigure}
\hspace{0.08\linewidth}%
\begin{subfigure}[t]{0.45\linewidth}
\includegraphics[]{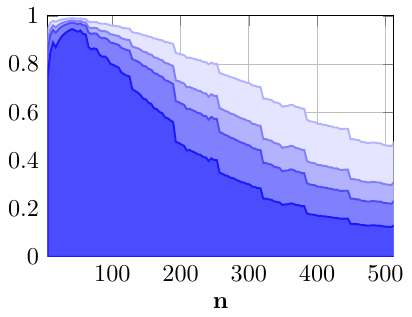}
\end{subfigure}
\caption{The fraction of time spent in copies/transpositions when computing the contraction $C_{mnp} = A_{mk} B_{pkn}$ using the conventional approach. Lines are shown with 1, 2, 3, and 6 total transpositions performed on either the input or output. (Left) CPU. (Right) GPU.}
\label{fig:TransRatio}
\end{figure}

\paragraph{Case study 1 }Consider $C_{mnp} = A_{mk} B_{nkp}$. The conventional approach presented in Section~\ref{sec:btas} results in an evaluation wherein one switches, by means of explicit copy operations, modes $n$ and $k$ in $B$ to produce $C_{mnp} = A_{mk} B_{knp}$, which is now of the form \eqref{eqn:BTASeq} and can be evaluated directly with a \GEMM. Alternatively, we observe that we may perform the computation without explicit copy by launching $\mathbf{p}$ individual \GEMMs.

\paragraph{Case study 2 }
Consider $C_{mnp} = A_{km} B_{pkn}$. The conventional approach presented in Section~\ref{sec:btas} results in an evaluation wherein we may require more than one transposition. For concreteness, we analyzed how \BTAS performs this contraction. We observed that \BTAS uses four explicit transpositions that results in the following algorithm:
\begin{compactenum}
\item Permute $A_{km}$ to $A_{mk}$.
\item Permute $B_{pkn}$ to $B_{kpn}$.
\item Permute $C_{mnp}$ to $C_{mpn}$.
\item Compute $C_{mpn} = \alpha A_{mk}B_{kpn} + \beta C_{mpn}$ with \GEMM.
\item Permute $C_{mpn}$ to $C_{mnp}$.
\end{compactenum}
Similarly, in the MATLAB Tensor Toolbox, the main idea is to reshape all tensors to matrices. For instance, in Case 2.4 in Table~\ref{tab:contract}, it reshapes $A_{km}$ to $A_{mk}$ and reshapes tensor $B_{pkn}$ to matrix $B_{k(pn)}$ with the first dimension as $\mathbf{k}$ and the second dimension as $\mathbf{p*n}$. Cyclops also uses index reordering methods for fully dense tensors. The reordering is avoided only in the more restrictive case of high-dimensional symmetric tensors. %However, these approaches are naive.

We note that some of the steps in the above approach can certainly be avoided with an improved algorithm that still implements the conventional approach. For example, Step 1 can be avoided by using a \GEMM that implicitly transposes the first matrix via a {\tt CblasTrans} parameter or equivalent in Step 4. Another optimization would be to avoid Step 3 altogether when $\beta = 0$. Other approaches require even fewer transposition steps. Ultimately, observe that we may perform the computation without explicit copy by performing $\mathbf{p}$ individual \GEMMs.

In Figure~\ref{fig:TransRatio}, we measure the cost of these explicit transpose operations in a representative tensor contraction on CPU and GPU. On the CPU we use \MKL's {\tt mkl\_somatcopy} and {\tt cblas\_sgemm}, and on the GPU we use \CUBLAS's {\tt cublasSgeam} and {\tt cublasSgemm} to perform each required matrix transposition and \GEMM respectively. Note that transposition primitives are not specified in \BLAS, but are vendor-specific \BLAS-like extensions provided to perform common transpose operations. For this reason, these optimized functions are not typically used in tensor libraries with, instead, custom transposition implementations taking their place. These custom implementations are likely not as optimized as the vendor implementations.

As we can see from Figure~\ref{fig:TransRatio}, on the CPU, almost 40\% of the time is used in copy and transpose, even when only a single mode transposition is performed. Clearly, with more transpose operations, the fraction is higher, requiring 60-80\% of the total time. This correlates well with data presented in~\cite{Li:2015:TensorMatrixMult} where it is reported that Tensor Toolbox takes approximately 70\% of the total time performing copies and transpositions in one algorithm. By avoiding these transpositions we may obtain 10x speedup on the GPU for small tensors with $\mathbf{n} \lesssim 100$, and more than 2x speedup on the CPU for almost all $\mathbf{n}$.

Although the fraction of time spent in transposition will asymptotically approach zero as $\mathbf{n}$ grows in both cases, the high bandwidth of the GPU allows the computation to dominate the communication much more quickly. 
Indeed, the reported maximum bandwidth of the K40c GPU is 288GB/sec and the dual-socket Xeon E5-2630 v3 CPU achieves 118GB/sec.
%Indeed, the reported maximum bandwidth of the K40c GPU is 288GB/sec and the peak performance is 4290GFlops/sec. While these are unattainable in practice, the ratio of performance to bandwidth is 14.9 flops/byte. For the dual-socket Xeon E5-2630 v3 CPU, the maximum bandwidth is 118GB/sec and peak performance is 1228GFlops/sec, for a ratio of 10.4 flops/byte. Thus, the GPU has a higher bandwidth relative to performance, which is able to mask the transposition operations more quickly. 

Additionally, that the gap between computational performance and communication performance continues to increase, so the cost of transposition is likely to increase in the future. Even now, especially for small tensor sizes, it is clear that the cost of performing explicit copies and transpositions is significant and should be avoided. %Later, we will discuss about the comparison between the explicit transpose strategy and our new strategy in Section~\ref{sec:explicittrans}. 

\vspace*{-6pt}
\subsection{Extended Notation}
\label{sec:notation2}

We would like to express evaluation strategies for tensor contractions succinctly, so we introduce additional notation.

In this paper, tensors are assumed to be stored in the \textit{column-major} format. In other words, the $i^\text{th}$ mode has a memory stride -- termed ``leading dimension" in \BLAS ~-- denoted {\tt ld<$i$>} with $\text{\tt ld<$0$>} = 1$. Using this notation, $A_{mnp}$ is stored as $A[m + n*\text{\tt ld<$1$>} + p*\text{\tt ld<$2$>}]$. Note that the common \textit{packed-storage} case is obtained when, for all $i$, we have $\text{\tt ld<$i$>} = \prod_{0 \leq k < i} \text{\tt dim<$k$>}$.
%\subsection{Extended Operations}

%\fixme{This needs to be presented differently. The purpose of these notations are that they map to the alternate evaluation techniques, not that this is how we're extending BLAS. By extending BLAS, we make these operations (performed at the innermost loop) as efficient as possible and avoid expensive copies/transpositions.}

%A standard \BLAS \GEMM kernel can perform operations like transpose and scaling on its matrix operands. To generalize this to accommodate tensor contractions, we propose to augment the set of formal parameters of the \GEMM kernel with a \textit{leading order} parameter to yield the \SBGEMM kernel. To this end,
We now formalize three operations that are used in tensor contraction evaluations.
%the following with an operation called batching, which is specified via the \textit{leading order}. We also propose two more operations applicable to tensors which can be achieved by appropriately specifying the dimension and the leading dimension arguments.
%\paragraph{Notation}
\begin{compactenum}
\item Batching: $[i]$ denotes that mode $i$ is {\em batched}, A batched mode is considered \textit{fixed}.
\item Flattening: $(ij)$ denotes that modes $i$ and $j$ are {\em flattened}, \ie, modes $i$ and $j$ are now considered together as a single mode. The combined mode $h = (ij)$ is considered \textit{free}.
\item Transpose: $A_{mn}^\top$ denotes a matrix {\em transpose}. Transposes may only be applied to tensors with exactly two free modes. %Note that this is same as the transpose in a standard \GEMM call except that we generalize it to allow for super-modes.
\end{compactenum}
The purpose of these notations is that they map directly to looped \BLAS calls and the appropriate evaluation can often be read directly from the notated expression. Next, we review some rules that the above notation must follow in order to obtain a well-formed evaluation expression.
%\paragraph{Rules}
\begin{compactenum}
\item A batched mode $[i]$ cannot be the first mode of any matrix term. That is, $A_{[m]nk}$ is not allowed. Batching in the first mode would cause the $\mathbf{m}$ resulting logical $\mathbf{n} \times \mathbf{k}$ matrices to be strided in both rows and columns and, therefore, cannot be used as a matrix in any \BLAS routine.
\item A flattening $(ij)$ requires that {\tt ld<$j$>} = {\tt ld<$i$>}$ \,${\tt dim<$i$>}. Unfortunately, the notation alone is therefore not sufficient to determine which modes may be flattened; it is contingent on the representation as well. In the common packed-storage case, however, this flattening condition is always true.
\item If a flattening operation occurs on the right side, it must occur on the left side with the same modes in the same order. For example, $C_{m(np)}$ = $A_{mk} B_{k(pn)}$ is not allowed.
\item Standard transposition rules apply: $C_{nm}^\top$ = $A_{mk} B_{kn}$ implies $C_{nm}$ = $B_{kn}^\top A_{mk}^\top$. However, modes may not be swapped under transposition. For example, $A_{mk}^\top$ can not be replaced with $A_{km}$.
\end{compactenum}

This notation allows us to quickly read off the intended extended \BLAS evaluation expression for arbitrary tensor contractions. See Table~\ref{tab:notationBLAS} for examples.

\begin{table*}[ht]
\centering
\scalebox{0.95}{
\setlength\extrarowheight{.2em}
\begin{tabular}{|c|r@{ }l|}
\hline
Contraction & \BLAS Evaluation & \\
\hline
$C_{m(np)} = A_{mk} B_{k(np)}$ & & \text{\GEMM('N','N', $\mathbf{m}$, $\mathbf{np}$, $\mathbf{k}$, 1, $A$, \texttt{lda<1>}, $B$, \texttt{ldb<1>}, 0, $C$, \texttt{ldc<1>});} \\
$C_{(mn)p} = B_{(mn)k} A_{pk}^\top$ & & \text{\GEMM('N','T', $\mathbf{mn}$, $\mathbf{p}$, $\mathbf{k}$, 1, $B$, $\texttt{ldb<1>}\cdot\texttt{ldb<2>}$, $A$, \texttt{lda<1>}, 0, $C$, $\texttt{ldc<1>}\cdot\texttt{ldc<2>}$);} \\
$C_{m[n]p} = A_{mk} B_{k[n]p}$ & \text{for $n$ in [0,$\mathbf{n}$)} & \text{\GEMM('N','N', $\mathbf{m}$, $\mathbf{p}$, $\mathbf{k}$, 1, $A$, \texttt{lda<1>}, $B+n\cdot\texttt{ldb<1>}$, \texttt{ldb<2>}, 0, $C+n\cdot\texttt{ldc<1>}$, \texttt{ldc<2>});} \\
$C_{mn[p]} = B_{k[p]m}^\top A_{kn}$ & \text{for $p$ in [0,$\mathbf{p}$)} & \text{\GEMM('T','N', $\mathbf{m}$, $\mathbf{n}$, $\mathbf{k}$, 1, $B+p\cdot\texttt{ldb<1>}$, \texttt{ldb<2>}, $A$, \texttt{lda<1>}, 0, $C+p\cdot\texttt{ldc<2>}$, \texttt{ldc<1>});} \\
$C_{[n]p} = B_{pk} A_{k[n]}$ & \text{for $n$ in [0,$\mathbf{n}$)} & \text{\GEMV('N', $\mathbf{p}$, $\mathbf{k}$, 1, $B$, \texttt{ldb<1>}, $A+n\cdot\texttt{lda<1>}$, 1, 0, $C+n$, \texttt{ldc<1>});} \\
\hline
\end{tabular}
}
\caption{Example mapping between tensor contractions with batched and flattened modes in our notation and the corresponding \BLAS expression evaluation. Note that the appropriate \BLAS primitive, transposition, matrix pointer, and leading dimension parameters to \GEMM can be read off directly from the notation.}
\label{tab:notationBLAS}
\end{table*}

\vspace*{-5pt}
\subsection{\BGEMM}
\begin{figure}[t]
\centering
\includegraphics[]{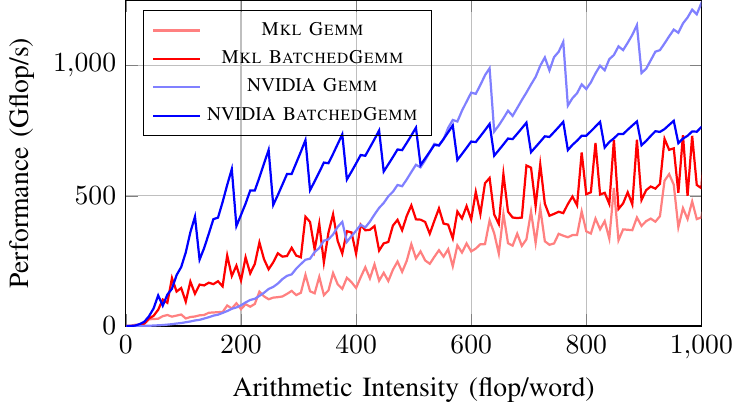}
\caption{The arithmetic intensity of computing $\mathbf{n}$ \GEMMs of size $\mathbf{n} \times \mathbf{n}$ versus the achieved performance on a K40c GPU and 16 cores (32 threads) of a dual socket CPU. }
\label{fig:intensity}
\end{figure}

Instead of relying on explicit mode transpositions, Peise \etal~\cite{Peise:2012:PMD,Peise:2015:TensorBLAS} considered mapping tensor contractions to \BLAS primitives directly -- enumerating all possible \BLAS primitives that could be used and their nesting within loops. Of course, the evaluation strategies that relied on level-3 \BLAS primitives (\GEMM) rather than level-2 primitives (\GEMV, \GER) were much more efficient. This often resulted in the need for many small \GEMMs to be performed, which usually does not achieve ideal performance.

The need to compute many small \GEMMs has not gone unnoticed by the leading implementations of \BLAS. NVIDIA supplied the capability to multiply pairs of many small matrices in \CUBLAS v4.1 [CUDA Toolkit v4.1] via the function {\tt cublasXgemmBatched}. %where {\tt X = S|D|C|Z}. This provided an alternative to launching many {\tt cublasXgemm} kernels, offered a natural target for small-matrix optimizations, and a provides a more reliable way to distribute the computation across the GPU multiprocessors. 
Similarly, as of \MKL 11.3$\beta$, {\tt cblas\_Xgemm\_batch} is available with a similar interface and is also specifically optimized for small matrix sizes.

In Figure~\ref{fig:intensity}, we plot the achieved performance on CPU and GPU of these \BGEMM functions by evaluating $\mathbf{n}$ \GEMMs of size $\mathbf{n} \times \mathbf{n}$ using each strategy with \MKL 11.3.1 and \CUBLAS 7.5. Note that there are much higher performance in both cases when $\mathbf{n}$ is small. When $\mathbf{n}$ is large, there is clearly room for optimization in {\tt cublasSgemmBatched}.

Both of these interfaces are based on pointers to matrix pointers, which often require allocation and/or precomputation at the point-of-call. This makes them awkward to use in the context of tensor contractions where the strides between matrices are regular and the generality provided by these interfaces goes unused. %However, the performance improvement due to the small-matrix optimizations implemented within each is evident.
\vspace*{-6pt}
\subsection{\SBGEMM}
\label{sec:SBGEMM}
Building on the \BGEMM extensions to \BLAS, we propose \SBGEMM (Listing~\ref{lst:interface}) which offers a simplified interface for the constant-strided \BGEMM and more optimizations opportunities for implementors. The interface and reference implementation of \SBGEMM is provided in Listing~\ref{lst:interface}. The {\tt lda}, {\tt ldb}, {\tt ldc} parameters are the standard ``leading dimension" parameters that appear in level-3 \BLAS primitives and denote to the stride between columns of the matrix. We refer to the new {\tt loa}, {\tt lob}, {\tt loc} parameters as the ``leading order" parameters and denote the stride between matrices of the batch. %Note that it is not necessarily the case that ${\tt lda} < {\tt loa}$, etc.

There are a number of advantages to a \SBGEMM primitive. First, \SBGEMM is actually more restrictive than the \BGEMM that has already appeared in \MKL and \CUBLAS, but we argue that a \BGEMM with a constant stride between matrices is a common enough case to consider specializing for. By providing this interface, the common case with constant strides between matrices is not forced to perform allocations or precomputations as it currently must perform in order to use \BGEMM. Additionally, these extra restrictions provide additional knowledge of the memory layout of the computation and offers additional optimizations opportunities in SIMDization, prefetching, and tiling. In other words, the "batch-loop" in \SBGEMM now directly participates in the polyhedral computation as an affine for-loop. With the pointer-interface in \BGEMM, the "batch-loop" cannot fully participate in a polyhedral model of the computation and is certainly not a candidate for vectorization or cache blocking.

\begin{table*}[ht]
\centering
\scalebox{0.80}{
\setlength\extrarowheight{.2em}
\begin{tabular}{|@{}c@{}|@{}c@{}|c|c|c||@{}c@{}|@{}c@{}|c|c|}
\hline
\ Case \ & \ Contraction \ & Kernel1 & Kernel2 & Kernel3  & \ Case \ & \ Contraction \ & Kernel1 & Kernel2 \\
\hline
1.1 & $A_{mk} B_{knp}$ & $C_{m(np)} = A_{mk} B_{k(np)}$ & $C_{mn[p]} = A_{mk} B_{kn[p]}$ & $C_{m[n]p} = A_{mk} B_{k[n]p}$ & 4.1 & $A_{kn} B_{kmp}$ & $C_{mn[p]} = B_{km[p]}^\top A_{kn}$ &   \\ 
1.2 & $A_{mk} B_{kpn}$ & $C_{mn[p]} = A_{mk} B_{k[p]n}$ & $C_{m[n]p} = A_{mk} B_{kp[n]}$ & & 4.2 & $A_{kn} B_{kpm}$ & $C_{mn[p]} = B_{k[p]m}^\top A_{kn}$ &  \\ 
1.3 & $A_{mk} B_{nkp}$ & $C_{mn[p]} = A_{mk} B_{nk[p]}^\top$ &  &  & 4.3 & $A_{kn} B_{mkp}$ & $C_{mn[p]} = B_{mk[p]} A_{kn}$ & \\ 
1.4 & $A_{mk} B_{pkn}$ & $C_{m[n]p} = A_{mk} B_{pk[n]}^\top$ &  & & 4.4 & $A_{kn} B_{pkm}$ & $TRANS(A_{kn}^\top B_{pk[m]}^\top)$  &  $C_{[m][n]p} = B_{pk[m]} A_{k[n]}$  \\ 
1.5 & $A_{mk} B_{npk}$ & $C_{m(np)} = A_{mk} B_{(np)k}^\top$ & $C_{mn[p]} = A_{mk} B_{n[p]k}^\top$ &  & 4.5 & $A_{kn} B_{mpk}$ & $C_{mn[p]} = B_{m[p]k} A_{kn}$ &  \\
1.6 & $A_{mk} B_{pnk}$ & $C_{m[n]p} = A_{mk} B_{p[n]k}^\top$ &  &  & 4.6 & $A_{kn} B_{pmk}$ & $TRANS(A_{kn}^\top B_{p[m]k}^\top)$ & $C_{[m][n]p} = B_{p[m]k} A_{k[n]}$  \\ 
\hline
2.1 & $A_{km} B_{knp}$ & $C_{m(np)} = A_{km}^\top B_{k(np)}$ & $C_{mn[p]} = A_{km}^\top B_{kn[p]}$ & $C_{m[n]p} = A_{km}^\top B_{k[n]p}$  & 5.1 & $A_{pk} B_{kmn}$ & $C_{(mn)p} = B_{k(mn)}^\top A_{pk}^\top$ & $C_{m[n]p} = B_{km[n]}^\top A_{pk}^\top $  \\
2.2 & $A_{km} B_{kpn}$ & $C_{mn[p]} = A_{km}^\top B_{k[p]n}$ & $C_{m[n]p} = A_{km}^\top B_{kp[n]}$ &   & 5.2 & $A_{pk} B_{knm}$ & $C_{m[n]p} = B_{k[n]m}^\top A_{pk}^\top$ &  \\
2.3 & $A_{km} B_{nkp}$ & $C_{mn[p]} = A_{km}^\top B_{nk[p]}^\top$ &  &  & 5.3 & $A_{pk} B_{mkn}$ & $C_{m[n]p} = B_{mk[n]} A_{pk}^\top$ &  \\
2.4 & $A_{km} B_{pkn}$ & $C_{m[n]p} = A_{km}^\top B_{pk[n]}^\top$ & &  &  5.4 & $A_{pk} B_{nkm}$ & $TRANS(B_{nk[m]} A_{pk}^\top)$ & $C_{[m]n[p]} = B_{nk[m]} A_{[p]k}$ \\
2.5 & $A_{km} B_{npk}$ & $C_{m(np)} = A_{km}^\top B_{(np)k}^\top$ & $C_{mn[p]} = A_{km}^\top B_{n[p]k}^\top$ &  & 5.5 & $A_{pk} B_{mnk}$ & $C_{(mn)p} = B_{(mn)k} A_{pk}^\top$ & $C_{m[n]p} = B_{m[n]k} A_{pk}^\top$   \\
2.6 & $A_{km} B_{pnk}$ & $C_{m[n]p} = A_{km}^\top B_{p[n]k}^\top$ & & & 5.6 & $A_{pk} B_{nmk}$ & $TRANS(B_{n[m]k} A_{pk}^\top)$ & $C_{[m]n[p]} = B_{n[m]k} A_{[p]k}$  \\
\hline
3.1 & $A_{nk} B_{kmp}$ & $C_{mn[p]} = B_{km[p]}^\top A_{nk}^\top$ &  & & 6.1 & $A_{kp} B_{kmn}$ & $C_{(mn)p} = B_{k(mn)}^\top A_{kp}$ & $C_{m[n]p} = B_{km[n]}^\top A_{kp} $  \\
3.2 & $A_{nk} B_{kpm}$ & $C_{mn[p]} = B_{k[p]m}^\top A_{nk}^\top$ &  &  & 6.2 & $A_{kp} B_{knm}$ & $C_{m[n]p} = B_{k[n]m}^\top A_{kp}$ &  \\ 
3.3 & $A_{nk} B_{mkp}$ & $C_{mn[p]} = B_{mk[p]} A_{nk}^\top$ &  & &  6.3 & $A_{kp} B_{mkn}$ & $C_{m[n]p} = B_{mk[n]} A_{kp}$ &  \\ 
3.4 & $A_{nk} B_{pkm}$ & $TRANS(A_{nk} B_{pk[m]}^\top)$ & $C_{[m][n]p} = B_{pk[m]} A_{[n]k}$ &  &  6.4 & $A_{kp} B_{nkm}$ & $TRANS(B_{nk[m]} A_{kp})$ & $C_{[m]n[p]} = B_{nk[m]} A_{k[p]}$  \\ 
3.5 & $A_{nk} B_{mpk}$ & $C_{mn[p]} = B_{m[p]k} A_{nk}^\top$ &  &  & 6.5 & $A_{kp} B_{mnk}$ & $C_{(mn)p} = B_{(mn)k} A_{kp}$ & $C_{m[n]p} = B_{m[n]k} A_{kp}$  \\ 
3.6 & $A_{nk} B_{pmk}$ & $TRANS(A_{nk} B_{p[m]k}^\top)$ & $C_{[m][n]p} = B_{p[m]k} A_{[n]k}$ &  & 6.6 & $A_{kp} B_{nmk}$ & $TRANS(B_{n[m]k} A_{kp})$ & $C_{[m]n[p]} = B_{n[m]k} A_{k[p]}$ \\
\hline

\end{tabular}
}
\caption{List of 36 possible single mode contraction operations between a second-order tensor and a third-order tensor and possible mappings to Level-3 \BLAS routines. Note that 8 cases may be performed with \GEMM, 28 cases may be performed with \SBGEMM, and 8 cases remain exceptional.}
\label{tab:contract}
\end{table*}

In Table~\ref{tab:contract}, we have enumerated all unique single-mode contractions between a second-order and third-order tensor using the notation from Section~\ref{sec:notation2}. All but 8 contractions can be computed with only a single call to \SBGEMM.

%Following, when we refer to \BGEMM we typically mean \SBGEMM unless otherwise noted.
\vspace*{-6pt}
\subsection{Exceptional Cases}
\label{sec:exceptionaleval}

The eight exceptional cases in Table~\ref{tab:contract} -- Cases 3.4, 3.6, 4.4, 4.6, 5.4, 5.6, 6.4, and 6.6 -- occur when batching forces the evaluation to either be a \BGEMV or violate the no-first-mode rule.

This can be resolved by making an extension to the operation parameters allowed for \BGEMM. Typically, the available operation parameters are ``normal", ``transpose", ``conjugate", and ``Hermitian". To account for the exceptional cases, ``extended X" could be added to allow violations of the no-first-mode rule and consider all three modes involved in the batching simultaneously.

For example, Case 3.6 and 6.4 could then be written
\begin{align*}
C_{mn[p]} = B_{[p]mk} \, A_{nk}^\top  && C_{m[n]p} = B_{[n]km}^\top \, A_{kp} 
\end{align*}
and evaluated via

\vspace{0.75em}
\hspace{-1em}
\begin{minipage}[b]{0.45\columnwidth}
{\scriptsize
\begin{verbatim}
sb_gemm(OP_EX_N, OP_T,
        M, N, K,
        1,
        B, ldb<1>, ldb<2>,
        A, lda<1>, 0,
        0,
        C, ldc<1>, ldc<2>,
        P); 
\end{verbatim}
}
\end{minipage}
\begin{minipage}[b]{0.45\columnwidth}
{\scriptsize
\begin{verbatim}
sb_gemm(OP_EX_T, OP_N,
        M, P, K,
        1,
        B, ldb<1>, ldb<2>,
        A, lda<1>, 0,
        0,
        C, ldc<2>, ldc<1>,
        N); 
\end{verbatim}
}
\end{minipage}

When the extended operation is passed, it is known that batching is in the first mode of the input which always has leading dimension 1. Thus, the leading order parameter to {\tt sb\_gemm} contains no information. Instead, leading dimensions of the other two modes in row-column order of the batched matrix are passed as the leading dimension and leading order parameters.

The implementation of a computation like this is expected to perform a ``3D'' tiling of $B$ into cache in order to efficiently contract with the standard 2D cache tiling of $A$.

%Note that this definition of {\tt sb\_gemm} with the extended operation parameters is a superset of \BLIS's \GEMM algorithm, which can be recovered with a batch size of one. \fixme{Check? The \BLIS output parameters might make it more general...}
\vspace*{-6pt}
\subsection{Generalization}
%\subsection{Generalization to Single-mode Contractions with Tensors of Arbitrary Order}

In this section, we explain the generality of our approach and how it can be easily applied and extended to single-mode contractions involving tensors of arbitrary order.

Consider an arbitrary single-mode tensor contraction of the form~\eqref{eqn:GETM}. It is straightforward to see by simple counting that the number of unique contractions is $[(|\mathcal{A}| + |\mathcal{B}| - 2)!] \cdot |\mathcal{A}| \cdot |\mathcal{B}|$. We note that Table~\ref{tab:contract} is obtained with $|\mathcal{A}|=2$ and $|\mathcal{B}|=3$. Of these contractions, all of them may be performed without explicit mode transpositions by nesting the \BGEMM operations.

We observe that some single-mode contractions of two tensors of arbitrary order can be evaluated by batching on different modes with the \BGEMM operations. For example, consider $C_{mn[p][q]} = A_{mk[p]} B_{nk[q]}$ wherein we can batch in either $p$ and $q$. We prefer to choose the mode with the larger dimension for the \BGEMM batching loop over the other (nested batching). %This does not negate the need for the exceptional evaluations, which must also be considered for nested batching.

The nested-batching strategy in Listing~\ref{code:nest2} is general and extends to any two tensors of any order. Algorithms and heuristics for choosing the looped, batched, and \GEMM-ed modes are provided in Section~\ref{sec:heuristic}.

\begin{cpp}[caption={Interface and reference implementation of BLAS-like strided batched \GEMM.}, label=lst:interface]
// C_p = alpha*opA(A_p)*opB(B_p) + beta*C_p
sb_gemm(op_type opA, op_type opB, 
        int m, int n, int k,
        T alpha, 
        const T* A, int lda, int loa,
        const T* B, int ldb, int lob,
        T beta,
        T* C, int ldc, int loc,
        int batch_size) 
{
  // EXPOSITION ONLY
  for (int p = 0; p < batch_size; ++p)
    gemm(opA, opB, 
         m, n, k,
         alpha,
         A + p*loa, lda,
         B + p*lob, ldb,
         beta,
         C + p*loc, ldc);
}
\end{cpp}
\vspace*{-5pt}
\begin{cpp}[caption={Nested batching.}, label=code:nest2]
for (int q = 0; q < Q; ++q)
  sb_gemm(OP_N, OP_T,
          M, N, K,
          1,
          A, lda<1>, lda<2>,
          B+q*ldb<2>, ldb<1>, 0,
          0,
          C+q*ldc<3>, ldc<1>, ldc<2>,
          P);
\end{cpp}

%=======================================================================
% eof

%
%%=======================================================================
\vspace*{-6pt}
\section{Results and Discussion}
\label{sec:results}
In this section, we benchmark varying evaluation strategies in order to define heuristics for computing general tensor contractions without copy or transposition. Additionally, we demonstrate the feasibility of the extended transpose parameter for exceptional case evaluations.

All performance measurements are performed on a heterogeneous CPU-GPU system with a dual-socket Intel Xeon E5-2630 v3 2.4GHz processor %(Haswell-EP) 
and an NVIDIA K40c GPU. Each CPU socket has 8 cores and 16 threads with an $8 \times 256$KB L2 cache and a 20MB L3 cache. The K40c has 2880 streaming cores distributed across 15 multiprocessors operating at 0.75GHz and a 1.5MB L2 cache.

All data used are randomized dense matrices. To eliminate noise from parallel competition of multi-sockets, all CPU results are generated from serial runs (one core, one thread).
\vspace{-5pt}
\vspace{-6pt}
\subsection{Conventional Evaluation}
\label{sec:explicittrans}

\begin{figure}[t]
\centering
\includegraphics[]{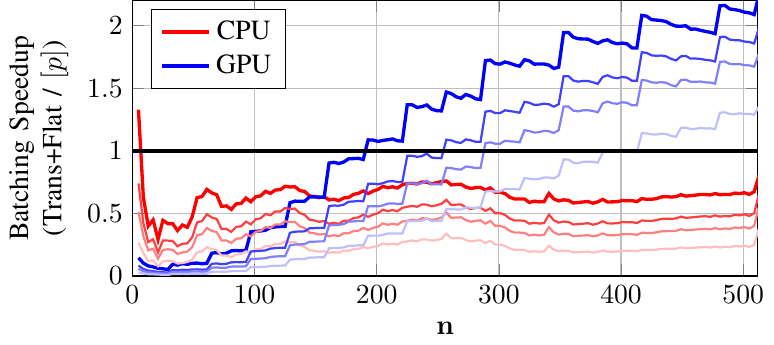}
\caption{Performance ratio between the conventional approach with $\kappa$ mode transpositions over a \BGEMM in $[p]$ for Case 1.3. For color from deep to light, $\kappa = 1,2,3,6$. Performance on CPU using \MKL's {\tt mkl\_somatcopy}, {\tt cblas\_sgemm}, and {\tt cblas\_sgemm\_batch}. Performance on GPU using \CUBLAS's {\tt cublasSgeam}, {\tt cublasSgemm}, and our modified {\tt cublasSgemmBatched}.}
\label{fig:Transflat}
\end{figure}

We further motivate the use of \SBGEMM evaluations by plotting the speedup of the conventional approach -- transpositions until a single \GEMM can be called -- over a single \SBGEMM call in evaluation of Case 1.3 from Table~\ref{tab:contract} for tensors of size $\v{n} \times \v{n} \times \v{n}$. Figure~\ref{fig:Transflat} shows that \SBGEMM is significantly faster than performing even a single transposition followed by a flattened \GEMM, especially for small matrices. Here, a single transposition means $\v{n}$ calls to {\tt mkl\_somatcopy} on CPU or {\tt cublasSgeam} on GPU in order to fully exchange two modes. The dark lines include only a single transposition and the lighter lines include 2, 3, and 6 transpositions.

On CPU, the \SBGEMM evaluation outperforms the conventional approach for all $\v{n} < 512$. On GPU, the benefit from performing a single flattened \GEMM eventually outweighs the cost of performing the transposition and for $\v{n} \gtrsim 200$ the conventional approach achieves a speedup over the \SBGEMM. This speaks to the highly optimized \GEMM in \CUBLAS and that, perhaps, additional optimization gains from \CUBLAS's \BGEMM may be available.

%we calculate the total time used with a single \BGEMM strategy over the total time used with 1 time transpose with flatten GEMM strategy for Case 1.3. The lighter color trails are estimations for the speedup with multiple transposes which corresponds to a BTAS-like strategy. The curves with the color varying from deep to light correspond to 1,2,3,6 times transposes with BTAS-like strategy.  As we can see from the figure, with tensor size less than 200, the speedup value is always less than 1. It shows contraction using BatchedGEMM strategy is faster than the one using BTAS-like strategy. In other words, copy/transpose time is slowing down the whole computation. After tensor size increase to more than 200, for GPU result, the Trans+Flat strategy is showing better performance due to excellent cublas implementations in GPU as it only need to do one GEMM computation while BatchedGEMM need paralleled computation and the communication time in it can't be omitted. In both CPU and GPU results, it's obvious to see that more transpose operations will reduce the efficiency of the BTAS-like strategy.
\vspace{-5pt}
\subsection{Extended BLAS Evaluation}

In this section, we %directly 
compare evaluation strategies given the extended \BLAS kernels. On GPU, the \SBGEMM interface is provided by modifying {\tt cublasSgemmBatched} from \CUBLAS 7.5. On CPU, the \SBGEMM interface is implemented in serial with looped calls to {\tt cblas\_sgemm} from \MKL 11.2. Both implementations thereby avoid additional allocation and/or precomputation at the call site. The serial execution on CPU emphasizes the cache effects discussed in the following sections.

\subsubsection{Flattening}

Cases 1.1, 1.5, and 6.1 can be evaluated without explicit transpositions with either a single flattened \GEMM or a single \BGEMM. We expect the flattened \GEMM evaluation to outperform the \BGEMM evaluation due to the optimization level of existing \GEMMs over that of the recently emerging \BGEMM functions.

\begin{figure}[t!]
\centering
\begin{subfigure}[t]{0.45\linewidth}
\includegraphics[]{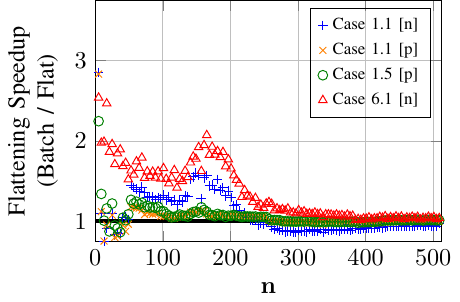}
\end{subfigure} 
\hspace{0.05\linewidth}%
\begin{subfigure}[t]{0.45\linewidth}
\includegraphics[]{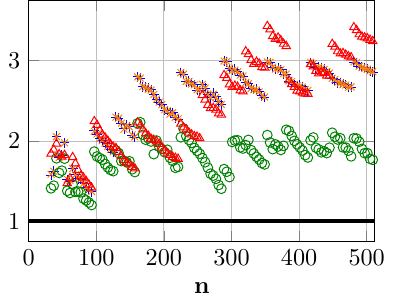}
\end{subfigure}
\caption{Performance ratio for a \BGEMM over a flattened \GEMM in evaluation of Cases 1.1, 1.5, and 6.1. (Left) CPU. (Right) GPU.}
\label{fig:FlatvBatch}
\end{figure}

In Figure~\ref{fig:FlatvBatch}, we plot the speedup achieved by using a flattened \GEMM evaluation over a \SBGEMM evaluation. 
%Note that in the following evaluation, the speedup of method A over method B is calculated by using ratio of B's total computation time over A's total time. 
In Figure~\ref{fig:FlatvBatch}, the speedup is greater than one when FlattenedGEMM is faster than the \SBGEMM. Clearly, most of the time, flattened \GEMM is faster. Furthermore, we note the \CUBLAS implementation of \SBGEMM is a great candidate for optimization as it appears to be significantly underperforming with respect to \GEMM.

We also note the dependence of the performance on the shape of the flattened \GEMM and the mode of the \SBGEMM. On CPU, we find that the major determining factor in performance is the batching mode of the output. That is, the \SBGEMM evaluation performs best when batched in the third mode of $C$ -- in Case 1.5 $[p]$ and 1.1 $[p]$. On GPU, the output batching mode makes no difference. It is unclear why the batched evaluation performs so well on Case 1.5 $[p]$. 

%Case 1.1 and 1.5 yields a \GEMM between a square matrix and a short-wide matrix while Case 6.1 yields a \GEMM between a tall-skinny matrix and a square matrix.

%show the speedup of due to using a flattened GEMM over a BatchedGEMM for Case 1.1, 1.5 and 6.1. If the speedup value greater than one, it means flattened GEMM is faster than BatchedGEMM. As we can see, the speedup is significant for small GEMM in CPU. Thus, when doing small tensor contraction, it's better to use flatten methods,while with large tensor size, we can choose BatchedGEMM method. The advantage of flattened GEMM is bigger in GPU setting.  Also, we notice that there is a difference in performance when the shape of the flattened matrix is different, specifically: short-wide shape versus tall-narrow shape. In both Case 1.1 and 6.1, the flattened matrices are short and wide while in Case 1.5, the flattened matrix is tall and narrow. From Figure~\ref{fig:FlatvBatch}, the former shape gives more advantages for the flattened GEMM strategy.

\subsubsection{Batching}
\label{sec:batching}

In this section, we attempt to quantify the performance gain by batching in the last mode versus an earlier mode and whether the input tensor or the output tensor should be prioritized for this optimization. 

\begin{figure}[t!]
\centering
\begin{subfigure}[t]{0.45\linewidth}
\includegraphics[]{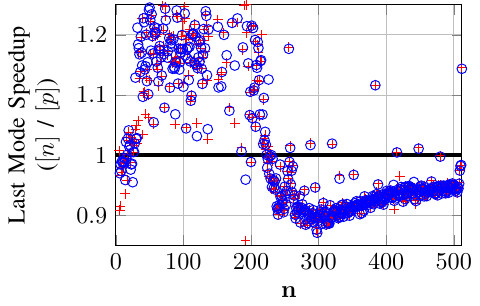}
\end{subfigure}
\hspace{0.08\linewidth}%
\begin{subfigure}[t]{0.45\linewidth}
\includegraphics[]{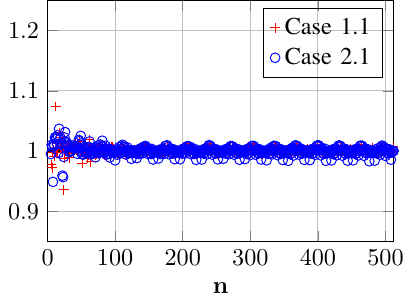}
\end{subfigure}
\caption{Speedup obtained from batching in the last mode, $[p]$, rather than the middle mode, $[n]$, for Cases 1.1 and 2.1. (Left) CPU. (Right) GPU.}
\label{fig:BatchNvP}
\end{figure}

Case 1.1 and 2.1 can both be batched in the second ($[n]$) or third ($[p]$) mode. In Figure~\ref{fig:BatchNvP}, we plot the speedup in performing the \BGEMM in $[p]$ over performing it in $[n]$. When the size of the tensor is small, $\mathbf{n} \lesssim 256$, batching in the third mode is advantageous and can result in up to 1.25x speedup on CPU. When $\mathbf{n} \gtrsim 256$, it is approximately 1.1x faster to batch in the second mode rather than the third. We expect this is an effect of the 256KB L1 cache, which would house the contiguous $B_{kn}$ submatrix for each $p$ when $\mathbf{n} \lesssim 256$. Beyond that size both batching strategies will have forced cache misses within each \GEMM, but by batching in the middle mode more data is shared between individual \GEMMs. %\fixme{Revisit again...}

On GPU, we see no discernible preference in the choice of batching mode. The GPU has a much less sophisticated memory system with no prefetcher and the performance difference is primarily determined by the number of global memory transactions issued. When $\mathbf{n} \geq 32$, the coalescing width is reached so nearly the same number of transactions will be issued in each case -- with small differences caused by alignment. We confirmed this by profiling the number of global memory reads and writes issued by each kernel and verifying that they correlate with the small differences in performance observed.

\begin{figure}[t!]
\centering
\begin{subfigure}[t]{0.45\linewidth}
\includegraphics[]{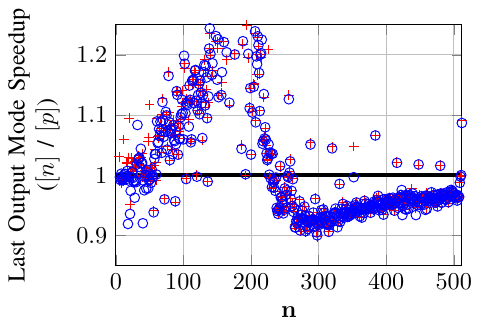}
\end{subfigure}
\hspace{0.08\linewidth}%
\begin{subfigure}[t]{0.45\linewidth}
\includegraphics[]{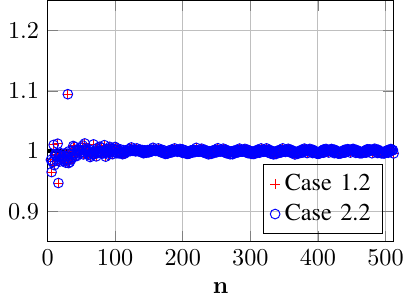}
\end{subfigure}
\caption{Speedup obtained from batching in the last output mode, $[p]$, rather than the middle output mode, $[n]$, for Cases 1.2 and 2.2. (Left) CPU. (Right) GPU.}
\label{fig:BatchIvO}
\end{figure}

Additionally, we consider the mixed-mode batching evaluations to determine if the input or output array is the primary determination of batching performance. In Figure~\ref{fig:BatchIvO}, we plot the speedup in performing \SBGEMM in the last mode of the output but the middle mode of the input, $[p]$, over performing it in the middle mode of the output and the last mode of the input, $[n]$, for Cases 1.2 and 2.2. The results are very similar to those of Figure~\ref{fig:BatchNvP} indicating that batching mode of the output tensor $C$ is more important than the batching mode of the input tensor $B$ on CPU. This is consistent with reference implementations of \GEMM which accumulate results directly into the output matrix.
\begin{comment}
Again, the GPU displays negligible difference in mixed-mode batching. The GPU accumulates results in local registers and shared memory before executing coalesced transactions to the output array. Again, only the alignment of these transactions affect the total number of coalesced reads and writes, but these differences are minute in these examples.

The implicit transposition inside the \SBGEMM -- flagged with the transpose parameter -- has no effect on performance since there are no major differences between Case 1.1 and 2.1 or between Case 1.2 or 2.2.
\end{comment}

\subsubsection{Exceptional Cases}
\label{sec:exceptionalresult}

\begin{figure}[ht]
\centering
\includegraphics[]{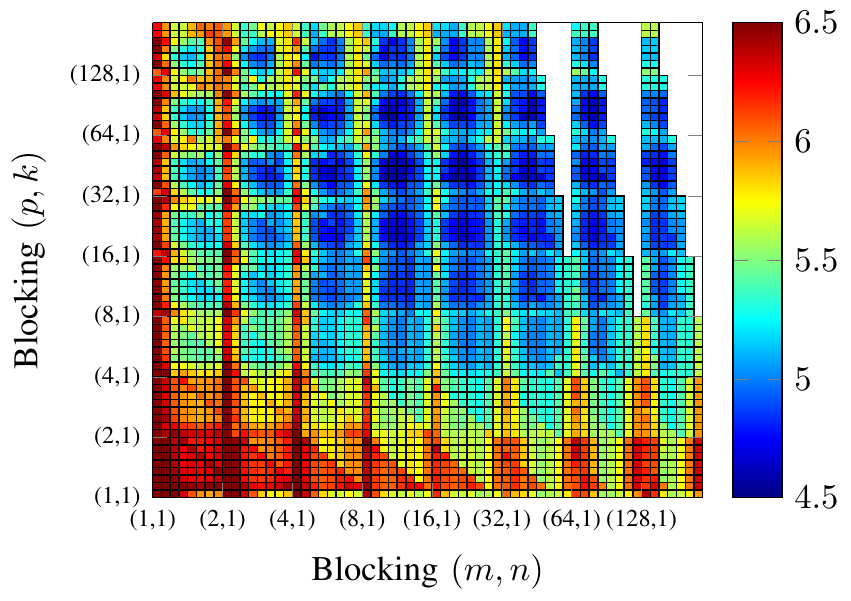}
\caption{GPU tiling parameter profile from PPCG on K40c for Case 6.4. Performance values are $\log_{10}([\mu sec])$ and tests performed for $\mathbf{m}=\mathbf{n}=\mathbf{k}=\mathbf{p}=256$. White indicates the run failed.}
\label{fig:gputiling}
\end{figure}

\begin{figure}[ht]
\centering
\includegraphics[]{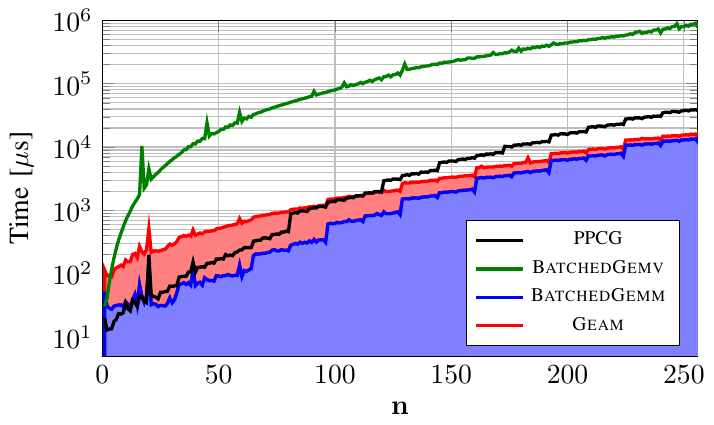}
\caption{Benchmark of three evaluation strategies for Case 6.4: A \BGEMV, a mode transposition followed by a \BGEMM, and an extended transpose kernel generated by PPCG.}
\label{fig:ppcgbench}
\end{figure}

In this section, we demonstrate the feasibility of evaluation strategies for the exceptional cases. 

The Polyhedral Parallel Code Generator (PPCG)~\cite{Verdoolaege:2013:PPCG} is a source-to-source compiler capable of generating CUDA kernels from nested control loops in C. We use PPCG to generate a CUDA kernel for exceptional Case 6.4 and compare its performance against other evaluation strategies.

First, Case 6.4 has four nested loops and PPCG accepts a tiling parameter for each. We search the parameter space $(m,n,p,k) \in [1,2,4,8,16,32,64,128]^4$ for the most efficient variant in Figure~\ref{fig:gputiling}. The kernels were generated with $\alpha = 1$ and $\beta = 0$ statically known as generated versions with dynamic $\alpha,\beta$ had significant branching and divergent overhead, whereas we are primarily interested in the access patterns and tiling. %In addition, many libraries, including \CUBLAS and \MKL, are optimized for the common case $\alpha=1$, $\beta = 0$ (e.g. skipping a load) so we do not feel the comparison is extraordinary. In contrast, the dimension of each mode is not provided statically.

The tiling parameters that result in the highest performance are $(16,4,32,4)$. Via inspection, we verify that the generated kernel is performing a 2D shared memory tiling for $A$, a ``3D'' shared memory tiling for $B$, and accumulating the $C$ results in registers. %Regardless, the kernel appears relatively unoptimized with potentially redundant or unnecessary bounds checks. No attempts were made to manually optimize and we suspect there is plenty of performance left on the table for expert implementations.

Using the $(16,4,32,4)$ kernel, we benchmark against two possible evaluation strategies: (1) A \BGEMV which requires no explicit transposition, and (2) A mode transposition in $k$ and $m$ followed by a \BGEMM in $[n]$. In Figure~\ref{fig:ppcgbench}, we show the execution time for each with the explicit transposition/\GEMM stacked to show their relative proportion in the two-step evaluation. The PPCG kernel outperforms the explicit transposition/\GEMM evaluation for small matrices and remains within a factor of 2-3x as $\mathbf{n}$ grows. We expect an expert implementation of the extended transpose parameter kernel would be able to close this gap and remain competitive with \BGEMM for all $\mathbf{n}$.
\vspace*{-6pt}
\subsection{Machine Learning Application}
\begin{figure}[ht]
\centering
\includegraphics[]{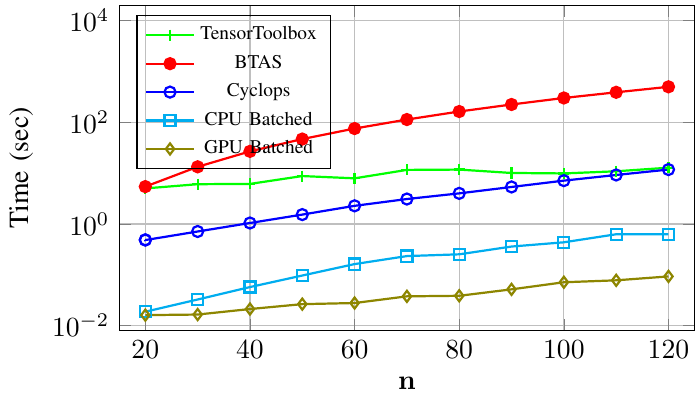}
\caption{Performance on Tucker decompostion.}
\label{fig:Tucker}
\end{figure}

\vspace*{-3pt}
In this section, we present the benchmarking results for the application that we discussed in Section~\ref{sec:ALS}.  For simulations on the CPU, we compare the performance on the Tucker decomposition using TensorToolbox, \BTAS, \CYCLOPS and our \SBGEMM. For simulations on the GPU, we don't have available GPU library to compare with, so we just evaluate our GPU implementation against \SBGEMM. %For all experiments, the input tensor is a three-mode tensor with identical dimension sizes ($\mathbf{n}$). 
We fix the number of iterations as $\mathcal{T} = 200$, set the core tensor size as $\mathbf{i} = \mathbf{j} = \mathbf{k} = 10$, and set the dimensions as $\mathbf{m} = \mathbf{n} = \mathbf{p} $. From Figure~\ref{fig:Tucker}, using our CPU \SBGEMM, we obtain more than 10 times speedup compared to \CYCLOPS /TensorToolbox and almost four orders of magnitude compared to \BTAS. Also, as expected, our GPU \SBGEMM confers further speedup.

\vspace*{-5pt}
\subsection{Evaluation Priorities}
\label{sec:heuristic}

Rather than attempt to model the algorithm and machine as in~\cite{Peise:2015:TensorBLAS,Napoli:2014:EfficientBLAS}, we simply provide evaluation guidelines based on the data provided. These are a number of heuristics that may be important in constructing the most efficient evaluation strategy.
\begin{compactenum}
\item Flatten modes whenever possible. A single large \GEMM is more efficient.% than a \BGEMM evaluating the same computation. This is not likely to change.
\item In the interest of performing the highest intensity computation within a \BGEMM, %and despite the data in Figure~\ref{fig:intensity}, 
we recommend performing the largest \GEMMs possible within a \BGEMM and batching in the mode with largest dimension.
\item Preferring to batch in the last mode versus earlier modes can depend on the input parameters and machine.
\end{compactenum}
We summarize these evaluation guidelines with pseudocode for performing a single-index tensor contraction %between two arbitrary-order tensors 
without copy or transposition in Algorithm~\ref{alg:Singlemodecontraction}.

\vspace*{-6pt}
\section{Conclusions and Future Work}
\label{sec:conclusion}
Our experience reveals that the emergence of \BGEMM provides significant computational advantages for multi-linear algebraic computations. The primitive allows us to push a larger high intensity computations to vendor-provided implementations. Leading implementations already provide \BGEMM on highly parallel machines. To simplify their use and provide additional optimization opportunities, we propose \SBGEMM and demonstrate its use for generalized tensor contractions. %Indeed, we have demonstrated the potential for in-place evaluation of all single-mode contractions between two tensors of arbitrary order with plausible extensions to these emerging \BLAS kernels. 
Calls to \SBGEMM have significant opportunity to perform at or near the performance of \GEMM and, by avoiding explicit transpositions or permutations of the data, accelerate these computations significantly.

Our improvement is most significant on small and moderate sized tensors. This is very important because in many applications, e.g. deep learning for training a recursive tensor network, we require evaluating a large number of tensor contractions of small sizes. % Also, we want to emphasize that our approach is not just about getting performance gains, but also having a simplified interface that can be then optimized over different platforms.

Although we focused on single-node performance, these evaluations may be used as building blocks for distributed memory implementations, which we intent to pursue as part of our future work. Further study into the optimized implementations, architecture-dependent implementations, and performance of the exceptional case kernels is warranted. More complicated contractions, such as multi-index contractions or sparse tensor algebra, also pose challenging problems. %We believe that our approach is only a first step in advancing the state-of-the-art multi-linear algebraic computations.

\begin{algorithm}[t]
   \algsetup{linenosize=\scriptsize}
  \footnotesize
\begin{algorithmic}[1]
    \STATE \textbf{In:} Tensor $A_\mathcal{A}$, $\mathcal{A} = [a_{1}, \dots, a_{M}]$, 
    \STATE \textbf{In:} Tensor $B_\mathcal{B}$, $\mathcal{B} = [b_{1}, \dots, b_{N}]$, $\mathcal{A} \cap \mathcal{B} = \{k\}$.
    \STATE \textbf{In, Out:} Tensor $C_\mathcal{C}$, $\mathcal{C} = [c_1,\ldots,c_{N+M-2}]$. WLOG, $c_1 \in \mathcal{A}$.
    \STATE Common substrings in $\mathcal{A}$, $\mathcal{B}$ and/or $\mathcal{C}$ for flattening candidates.
    %\STATE Common substrings in $\mathcal{B}$ and $\mathcal{C}$ for flattening candidates
    \STATE Relabel flattened modes
    \STATE Compute $\mathcal{P} = \{c_i \ | \ i \neq 1, c_i \not\equiv a_1, c_i \not\equiv b_1\}$
    %\STATE WLOG, let $c_1 \in \mathcal{A}$
    \IF{$\abs{\mathcal{C} \setminus \mathcal{P}} = \abs{\{c_1\}} = 1$.}
      \STATE [Case $C_{c_1\cdots} = A_{k\cdots c_1\cdots}B_{k\cdots}$]
      \STATE Let $c^* \in \mathcal{P}\setminus\mathcal{A}$ be index with max dimension
      \STATE Let $c^+ \in \mathcal{P}\setminus\{c_1,c^*\}$ be index with max dimension
      \STATE Nested in all $c_j \in P \setminus \{c^*,c^+\}$, \BGEMM in $c_1, c^*, k, [c^+]$
	\ELSIF{$\abs{\mathcal{C} \setminus \mathcal{P}} = \abs{\{c_1, c_b\}} = 2$}
	  \STATE [Case $C_{c_1\cdots c_b \cdots} = A_{k\cdots c_1\cdots}B_{c_b\cdots k \cdots}$]
	  \STATE Let $c^* \in \mathcal{P}$ be index with max dimension
	  \STATE Nested in all $c_j \in P \setminus \{c^*\}$, \BGEMM in $c_1, c_b, k, [c^*]$
	\ELSIF{$\abs{\mathcal{C} \setminus \mathcal{P}} = \abs{\{c_1, c_a\}} = 2$}
	  \STATE [Case $C_{c_1\cdots c_a \cdots} = A_{c_a\cdots c_1\cdots k \cdots}B_{k\cdots}$]
	  \STATE Let $c^* \in \mathcal{P}\setminus \mathcal{A}$ be index with max dimension
	  \STATE Nested in all $c_j \in P \setminus \{c^*\}$, Ex. \BGEMM in $c_1, c^*, k, [c_a]$
	\ELSIF{$\abs{\mathcal{C} \setminus \mathcal{P}} = \abs{\{c_1, c_a, c_b\}} = 3$}
	  \STATE [Case $C_{c_1\cdots c_a \cdots c_b \cdots} = A_{c_a\cdots c_1\cdots k \cdots}B_{c_b\cdots k \cdots}$]
	  \STATE Nested in all $c_j \in P$, Ex. \BGEMM in $c_1, c_b, k, [c_a]$
	\ENDIF
	\caption{Single-Mode Tensor Contraction}
    \label{alg:Singlemodecontraction}
\end{algorithmic}
\end{algorithm}

\vspace*{-10pt}
\section*{Acknowledgment}
The authors would like to thank Aparna Chandramowlishwaran for providing the computation resources and suggestions. Animashree Anandkumar is supported in part by Microsoft Faculty Fellowship, NSF Career Award CCF-1254106, ONR Award N00014-14-1-0665, ARO YIP Award W911NF-13-1-0084, and AFOSRYIP FA9550-15-1-0221. Yang Shi is supported by NSF Career Award CCF-1254106 and ONR Award N00014-15- 1-2737,Niranjan is supported by NSF BigData Award IIS-1251267 and ONR Award N00014-15-1-2737.



%\bibliographystyle{IEEEtran}
%\begin{thebibliography}{1}

\vspace*{-10pt}
\bibliographystyle{plain}
\bibliography{bare_conf}

\end{document}